\documentclass[conference]{IEEEtran}
\IEEEoverridecommandlockouts
\usepackage{hyperref}
\hypersetup{colorlinks=true,citecolor=blue}
\usepackage{cite}
\usepackage{amsmath,amssymb,amsfonts}
\usepackage{algorithmic}
\usepackage{graphicx}
\usepackage{textcomp}
\usepackage{xcolor}
\def\BibTeX{{\rm B\kern-.05em{\sc i\kern-.025em b}\kern-.08em
    T\kern-.1667em\lower.7ex\hbox{E}\kern-.125emX}}
\begin{document}

\title{`Complementarity' in paraxial and non-paraxial optical beams
   \thanks{Corresponding author: pprasanta@iiserkol.ac.in}}

\author{
    \IEEEauthorblockN{
        Abhinash Kumar Roy\IEEEauthorrefmark{1}\IEEEauthorrefmark{2}, 
        Nitish Kumar Chandra\IEEEauthorrefmark{1}\IEEEauthorrefmark{3},
        Soumik Mahanti\IEEEauthorrefmark{1}\IEEEauthorrefmark{4}, and 
        Prasanta K. Panigrahi\IEEEauthorrefmark{1}
    }
    \IEEEauthorblockA{
        \IEEEauthorrefmark{1}Department of Physical Sciences, Indian Institute of Science Education and Research Kolkata, Mohanpur-741246, West Bengal, India
    }
    \IEEEauthorblockA{
        \IEEEauthorrefmark{2} Department of Physical and Mathematical Sciences, Macquarie University, Sydney NSW, Australia
    }
    \IEEEauthorblockA{
        \IEEEauthorrefmark{3} Department of Informatics \& Networked Systems, University of Pittsburgh, Pittsburgh, PA 15213, USA
    }
    \IEEEauthorblockA{
        \IEEEauthorrefmark{4} S.N. Bose National Center for Basic Sciences, Block JD, Sector III, Salt Lake, Kolkata 700106, India
    }
 }

\maketitle

\begin{abstract}
    Establishing the correspondence of two dimensional paraxial and three dimensional non-paraxial optical beams with the qubit and qutrit systems respectively, we derive a complementary relation between Hilbert-Schmidt coherence, generalized predictability and linear entropy. The linear entropy, a measure of mixedness is shown to saturate the complementarity relation for mixed bi-partite states. For pure two qubit and qutrit systems, it quantifies the global entanglement and reduces the complementarity relation to the triality relation between coherence, predictability and entanglement. We analyze these relations in wedge-product formalism in order to investigate the innate geometry of the complex vector space. The derived complementary relations offer insights into our ability to manipulate and utilize quantum properties for practical advancements.
\end{abstract}


\section{Introduction}
Entanglement has played a significant role in advancing our understanding of multiparty quantum systems \cite{schrodinger1936probability, Erhard2020}. This  inseparability of a composite system is a quantum mechanical property, which is absent in classical systems. 
However, the analogous mathematical structure of vector spaces in classical optics \cite{born_wolf_bhatia_clemmow_gabor_stokes_taylor_wayman_wilcock_1999, wolf2007introduction} enables the realization of local entanglement \cite{Lorenzo:wiley:2020} in a single system arising out of the inseparability between its various degrees of freedom. This type of intra-system entanglement has been referred to as classical or non-quantum entanglement \cite{simon:prl:2010,Eberly,Toppel}, and it manifests in optical beams, when the polarization state and spatial degrees of freedom are inseparably entangled.

The similarity in mathematical structure has paved the way for leveraging classical entanglement in computational benefits, a domain previously considered exclusive to quantum entanglement. Numerous studies have successfully employed classical entanglement to implement quantum algorithms, highlighting its effectiveness in tasks like the Deutsch algorithm~\cite{PEREZGARCIA20151675} and quantum walk~\cite{PhysRevLett.110.263602}. Additionally, classical entanglement has proven practically useful in communication applications, covering tasks such as quantum channel tomography~\cite{Ndagano2017} and teleportation~\cite{Silva_2016}. In a recent work~\cite{You:21}, two detection methods were introduced to quantify the Schmidt number, a metric used for characterizing non-separable correlations between polarization direction and amplitude in a light field.

Beyond qubits, qutrit systems offer a substantial expansion of the accessible state space for information storage and processing \cite{wang2020qudits}. The three-level systems or qutrits as fundamental components in quantum information processing emerges as a compelling alternative to prevailing qubit-based architectures, presenting substantial computational advantages. This intrinsic attribute not only facilitates a reduction in circuit complexity but also simplifies experimental setups and has the potential to enhance the efficiency of various quantum algorithms \cite{gokhale2019asymptotic}. This progress marks a significant step toward the development of fully programmable quantum processors capable of achieving universal quantum computing \cite{PhysRevApplied.19.064024}.

Traditionally, the complementarity has been explored in the context of the wave-particle duality of quantum objects \cite{Bohr1928-BOHTQP,wooters:zurek:PRD:1979, GREENBERGER:yasin:PLA:1988}. Recent investigations have revealed the complementarity relations in optical systems having different non-separable degrees of freedom \cite{eberly:prl:2016, eberly:prr:2020, Zela:ol:2018}. One such complementarity exists between the degree of polarization and the entanglement measure, concurrence \cite{wooters:PRL:1998}. The degree of polarization is related to the coherence and predictability via Polarization Coherence Theorem \cite{eberly:optica:PCT}. This leads to the triality relation \cite{jacob:optcomm:2010, Basso_2020,tabish:ol:2021}, $P^{2}+C^{2}+\mathcal{E}^{2}=1$, connecting coherence ($C$), predictability ($P$) and entanglement ($\mathcal{E}$). These relations hold for paraxial electromagnetic field having planar wavefront, that are described by a $2\times2$ polarization coherence matrix. 

In optics, paraxial and non-paraxial beams offer distinct descriptions of light propagation~\cite{Nemoto:90,Hazarika_2019}. While paraxial calculations assume small angles for simplicity, this approach falls short with divergent or convergent beams and large numerical aperture systems. To accurately represent such scenarios, the non-paraxial regime incorporates higher-order variables in the wave equation. This becomes crucial for understanding intricate optical processes, particularly in cases where the paraxial approximation, like in self-focusing in nonlinear media, may prove insufficient. 

In this work, we investigate the complementarity for 3-D non paraxial electromagnetic fields, where the wavefronts are not necessarily planar, and described by a $3\times3$ polarization-coherence matrix. The extension of polarization-coherence matrix for this case requires a generalization of Stokes parameter, which from being four parameters in the two dimensional case will now have nine parameters  \cite{friberg:PRE:2002, wolf:optComm:2005, sheppard:PRA:2014}. Also, the expansion of the general state in terms of the Hermitian matrix requires the Gell-Mann matrices \cite{Gell:PhyRev:1962:} and the description of the non-separability of different degrees of freedom of
optical beams requires the extension of the concurrence to multipartite qudit systems, namely I-concurrence \cite{rungta:PRA:2001, bhaskara:QIP:2017}. 
We use the correspondence of a paraxial and non-paraxial electromagnetic field having two and three independent degrees of freedom with the qubit and qutrit systems respectively. We then derive a complementarity relation involving Hilbert-Schmidt coherence, generalized predictability and a measure of mixedness, linear entropy which saturates for mixed states. For pure bi-partite systems, the linear entropy takes the form of entanglement measure, concurrence and the complementarity relation reduces to the usual triality relation \cite{jacob:optcomm:2010}. 
\section{Preliminaries}
Here we briefly describe and motivate the terms that will appear in the complementarity relations we discuss in the following sections.

\textit{Predictability}: The generalized predictability is defined as~\cite{PhysRevA.105.032209},
\begin{equation} \label{pred}
    \mathcal{P}^{2} = \frac{2(n-1)}{n}\left[\sum_{i = 1}^{n}(\Phi_{ii})^{2} - \frac{2}{n-1}\sum_{i<j}\Phi_{ii}\Phi_{jj}\right]
\end{equation}
where, $n$ is the dimension of the polarization-coherence matrix $\Phi$. It is apparent that the predictability depends only on the diagonal terms of the matrix, i.e., on the fraction of energy across the orthogonal components of the field. The above definition of predictability is motivated by the following reasons. Firstly, it satisfies the boundary conditions i.e., zero when the fraction of energy across the orthogonal component of the field is equal and maximum when only one of the orthogonal component is non-zero. Secondly, with this predictability, the entanglement measure that appears in the complementary relations is the well established measure concurrence \cite{wooters:PRL:1998}, and its extension I-concurrence \cite{rungta:PRA:2001, bhaskara:QIP:2017}, for 2-D and 3-D pure bipartite fields respectively. Furthermore, this definition along with the Hilbert-Schmidt coherence extends the well known polarization-coherence theorem \cite{eberly:optica:PCT} for the 3-D case, which leads to the important result that, degree of polarization is given by predictability only when $\Phi$ is diagonal \cite{Qasimi:OL:2007}.

\textit{Linear entropy}: The linear entropy is given by,
\begin{equation}
    \mathcal{M}^{2}(\rho) = \frac{d}{d-1}(1-\operatorname{Tr}(\rho^{2})),
\end{equation}
where, $d$ is the dimension of the density matrix $\rho$. It is obtained by approximating von Neumann entropy and retaining its leading order term. This quantity ranges from $0$ to $1$, the former value is for the pure states and the latter value for the maximally mixed state. For two level system given by the reduced density matrix $\rho_{A}$, it becomes, $\mathcal{M}^{2}(\rho_{A}) = 2(1-\operatorname{Tr}(\rho_{A}^{2}))$.

\textit{Hilbert-Schmidt coherence}: The Hilbert-Schmidt coherence~\cite{OZAWA2000158,PhysRevA.103.042423} is defined in terms of off-diagonal elements as, 
\begin{equation}
    \mathcal{C}^{2} = 2\sum_{i\neq j}|\Phi_{ij}|^{2},
\end{equation}
where $\Phi$ represents the state.
\section{Complementarity in paraxial beams}
Electric field of a non-uniformly polarized beam of light with planar wavefront in the $x-y$ plane, can be described by,
\begin{equation} \label{el} 
   \textbf{E}(\textbf{r}) = a \textbf{e}_{x} \psi(\textbf{r}) + b \textbf{e}_{x} \phi(\textbf{r}) + c\textbf{e}_{y} \psi(\textbf{r}) + d \textbf{e}_{y} \phi(\textbf{r}),
\end{equation}
where, $a,b,c,d$ are complex coefficients and,  $\psi(\textbf{r})$ and $\phi(\textbf{r})$ are orthonormal functions. Depending on the coefficients, the polarization and spatial degrees of freedom (DOF) can be separable or entangled. To meaningfully characterize the amount of inseparability, we can find a correspondence of this form of non-uniformly polarized beam with four dimensional Hilbert space, $\mathcal{H} = \mathcal{H}_{p}\otimes\mathcal{H}_{s}$, where, $\mathcal{H}_{p} = \text{Span}\{\textbf{e}_{x}, \textbf{e}_{y}\}$ and $\mathcal{H}_{s} = \text{Span}\{\psi(\textbf{r}), \phi(\textbf{r})\}$, and following the prescription, $|0\rangle_{p} = \textbf{e}_{x}$, $|1\rangle_{p} = \textbf{e}_{y} $, $|0\rangle_{s} = \psi(\textbf{r})$, $|1\rangle_{s} = \phi(\textbf{r})$, the electric field in (\ref{el}) can be identified as \cite{Toppel},
\begin{equation} \label{qubit1}
    |\textbf{E}\rangle = a|00\rangle + b|01\rangle + c|10\rangle + d|11\rangle
\end{equation}
where, $|ij\rangle = |i\rangle_{p}\otimes |j\rangle_{s}$, and $|\textbf{E}\rangle $ is a vector in the Hilbert space $\mathcal{H}$. With this identification, one can use the standard techniques from quantum information to analyse the entanglement. We leverage parallels from the notion of post-measurement vectors \cite{bhaskara:QIP:2017, Roy2021,Mahanti_2023} and extend its application to optical beams. These vectors are defined as the projection of each quantum state of the subsystem onto the complete state. By employing these post-measurement vectors, the geometry of entanglement can be described in terms of area vectors. For instance, from configuration of the vectors, $\langle 0_{p}|\textbf{E}\rangle = a|0\rangle_{s} + b|1\rangle_{s}$ and $\langle 1_{p}|\textbf{E}\rangle = c|0\rangle_{s} + d|1\rangle_{s}$ we can find whether the polarization and spatial degree of freedom are separable or maximally entangled.
The separability of polarization and spatial DOF corresponds to the $\langle 0_{p}|\textbf{E}\rangle$ and $\langle 1_{p}|\textbf{E}\rangle$ being parallel in $\mathcal{H}_{s}$~\cite{bhaskara:QIP:2017}  i.e.,
\begin{equation*}
  \text{Separability} \iff  \frac{a}{b} =\frac{c}{d}
\end{equation*}
 whereas the orthogonality and equality of $\langle 0_{p}|\textbf{E}\rangle$ and $\langle 1_{p}|\textbf{E}\rangle$ correspond to the maximal entanglement, i.e., 
 \begin{equation*}
 \begin{aligned}
     \text{Maximal entanglement}\iff &ac^{*} + bd^{*} = 0 ~~\text{and},~ \\&|a|^{2} + |b|^{2} = |c|^{2} + |d|^{2}.
     \end{aligned}
 \end{equation*}
 The above description provides a geometrical perspective of separability and entanglement in terms of configuration of vectors in the Hilbert space and matches with the usual description in terms of reduced density matrix being a projector and maximally mixed for the respective scenarios.

The reduced density matrix corresponding to the polarization degree of freedom is known as polarization-coherence matrix, which for the pure state $|\textbf{E}\rangle$ is obtained as,
\begin{equation}
    \Phi = \begin{bmatrix} \langle 0_{p}|\textbf{E}\rangle^{\dagger}\langle 0_{p}|\textbf{E}\rangle & \langle 1_{p}|\textbf{E}\rangle^{\dagger}\langle 0_{p}|\textbf{E}\rangle\\ \langle 0_{p}|\textbf{E}\rangle^{\dagger}\langle 1_{p}|\textbf{E}\rangle & \langle 1_{p}|\textbf{E}\rangle^{\dagger}\langle 1_{p}|\textbf{E}\rangle \end{bmatrix}.
\end{equation}
We note that $\Phi$ is a Hermitian matrix and, $\operatorname{Tr}(\Phi)$ corresponds to the intensity $\langle\textbf{E}|\textbf{E}\rangle$ of the beam. Moreover, since the above is obtained for a pure state,  $\operatorname{det}(\Phi) = 0$. In general, polarization-coherence matrix for a beam with planar wavefront is a $2\times 2$ Hermitian, positive semi-definite matrix is,
\begin{equation}\label{polcoh2}
    \Phi = \frac{1}{2}(I + \sum_{i = 1}^{3}S_{i}\sigma_{i}),
\end{equation}
where, $S_{i} = \langle\sigma_{i}\rangle = \operatorname{Tr}(\Phi\sigma_{i})$ are the Stokes parameters and $\{\sigma_{1}, \sigma_{2}, \sigma_{3}\}$ are the Pauli matrices. As a consequence of positive semi-definiteness of $\Phi$, we obtain $S_{1}^{2} + S_{2}^{2} + S_{3}^{2} \leq 1$. Therefore, set of all polarization-coherence matrix of a beam with planar wavefront can be represented by a solid ball of unit radius (Poincare sphere) with the center representing the unpolarized beam (maximally mixed state) and the points on the surface representing the polarized beams (pure states).


For paraxial beams given by (\ref{polcoh2}), the predictability obtained is the difference in intensity corresponding to $|0\rangle_{p}$ and $|1\rangle_{p}$,   
\begin{equation}
    \mathcal{P}^{2} = \left(\Phi_{11} - \Phi_{22}\right)^{2} = S_{3}^{2},
\end{equation}
equivalently, it is the square of expectation value of the observable $\sigma_{z}$.
Hilbert-Schmidt coherence for paraxial beams in the basis $\{|i\rangle_{p}\}$ is obtained as, 
\begin{equation}
    \mathcal{C}^{2} = 2\sum_{i\neq j }|\Phi_{ij}|^{2}
\end{equation}
where $i, j = 1,2$. In terms of the observable, it is obtained from the expectation values of $\sigma_{x}$ and $\sigma_{y}$ as,
\begin{equation}
    \mathcal{C}^{2} = \langle\sigma_{x}\rangle^{2} + \langle\sigma_{y}\rangle^{2}.
\end{equation}
For the pure states, predictability and coherence obey the well known duality relation \cite{pati_tabish:PRA:2015},
\begin{equation}
\begin{aligned}
    \mathcal{P}^{2} + \mathcal{C}^{2} &=  \langle\sigma_{x}\rangle^{2} + \langle\sigma_{y}\rangle^{2} + \langle\sigma_{z}\rangle^{2} \\
    &= S_{1}^{2} + S_{2}^{2} + S_{3}^{2} = 1,
    \end{aligned}
\end{equation}
which reflects the complementary character of these properties. When $\mathcal{P}$ is maximum, i.e., when the beam is in the state $|0\rangle_{p}$ or $|1\rangle_{p}$, the coherence is zero, whereas when the predictability is zero i.e., when $|a|^{2} + |b|^{2} = |c|^{2} + |d|^{2}$ for the beam (\ref{el}), the coherence obtained is maximum.

When we deal with a bipartite system, with each party representing a two level system, state of each party is described by the reduced density matrix thus obtained after tracing out the other. In general, the reduced density matrix obtained is not pure, i.e., cannot be represented by a particular state vector, rather an ensemble.  In classical optics, the reduced density matrix corresponding to the polarization DOF represents the polarization-coherence matrix, $P^{2} = \sum_{i} S_{i}^{2}$, also known as degree of polarization of the beam. This realization trivially leads to the polarization-coherence theorem, $P^{2} = \mathcal{P}^{2} + \mathcal{C}^{2}$  \cite{eberly:optica:PCT}. Since the polarization coherence matrix is mixed in general, the duality relation becomes, $\mathcal{P}^{2} + \mathcal{C}^{2} \leq 1$. 

Pure bipartite systems are described by a state vector of the form (\ref{qubit1}). For fields given by (\ref{qubit1}), if the polarization and spatial DOF is separable, the resulting polarization-coherence matrix represents a polarized state and the duality relation saturates. However, if both DOF are inseparable, the resulting polarization coherence matrix represents a partially polarized state. It has been shown that the amount of entanglement quantified by the concurrence \cite{wooters:PRL:1998}, which for the state (\ref{qubit1}) is given by, $\mathcal{E} = 2|ad-bc|$ obeys a triality relation with predictability and coherence, $\mathcal{E}^{2} + \mathcal{P}^{2} + \mathcal{C}^{2} = 1$ \cite{jacob:optcomm:2010}. Therefore, the amount of inseparability of polarization and spatial DOF bounds the degree of polarization of the beam \cite{eberly:prl:2016, Zela:ol:2018}. If the beam itself is mixed the above triality relation becomes an inequality \cite{Qasimi:josaA:2020}. 

When the bipartite system itself is mixed, concurrence does not remain a bonafide measure of mixedness of the reduced density matrices, therefore fails to saturate the complementary relation for the subsystems. Concurrence provides a necessary and sufficient condition for the separability of the density matrix, that is, whether the density matrix can be obtained from an ensemble containing only pure states. Therefore, even when the concurrence is zero, the reduced density matrix can be mixed, and hence $\mathcal{P}^{2} + \mathcal{C}^{2} <1$. We show that the linear entropy of mixedness saturates the complementary relation in general, and for the pure two qubit state, this exactly quantifies the global entanglement concurrence. Suppose the reduced density matrix thus obtained is of the form $\Phi = (I + \textbf{S}_{A}\cdot{\sigma})/2$, one obtains,
\begin{equation}
    \mathcal{M}^{2}_{A} + \mathcal{C}_{A}^{2} + \mathcal{P}_{A}^{2} = 2\left[1- \operatorname{Tr}\left(\frac{I(1 +|\textbf{S}_{A}|^{2})}{4}\right)\right] + |\textbf{S}_{A}|^{2} = 1,
\end{equation}
 where, $\mathcal{M}_{A} = \mathcal{M}(\Phi)$. 
 
We observe that the amount of mixedness in a two level system bounds the total amount of local information it can possess in the form of coherence and predictability. It is worth noting that the linear entropy of mixedness for a two level system reduces to, $ \mathcal{M}^{2}(\Phi) = 2(1-\operatorname{Tr}((\Phi)^{2})) = 4\operatorname{det}(\Phi)$, which for a pure bi-partite state exactly quantifies the entanglement between the subsystem, however, such is not the case for the mixed bipartite state. For the pure bipartite systems, the mixedness of the individual subsystem arises only due to the entanglement between two party. Therefore, the triality relation involving mixedness reduces to the triality relation involving the concurrence for the pure-bipartite system.
\section{Complementarity in non-paraxial beams}
Electromagnetic waves with arbitrary wavefront is given by a $3\times3$ Hermitian, positive semi-definite polarization-coherence matrix, therefore isomorphic to a density matrix of a three level (qutrit) system \cite{sheppard:PRA:2014}.  
The general form of polarization-coherence matrix requires the use of Gell-Mann matrices $\{\lambda_{i}\}$ given by \cite{Gell:PhyRev:1962:},
\begin{align*}
    \lambda_{1} &= \begin{pmatrix} 0 & 1 & 0 \\ 1 & 0 & 0 \\ 0 & 0 & 0 \end{pmatrix} \quad \lambda_{2} = \begin{pmatrix} 0 & -i & 0 \\ i & 0 & 0 \\ 0 & 0 & 0 \end{pmatrix} \quad \lambda_{3} = \begin{pmatrix} 1 & 0 & 0 \\ 0 & -1 & 0 \\ 0 & 0 & 0 \end{pmatrix} \\
    \lambda_{4} &= \begin{pmatrix} 0 & 0 & 1 \\ 0 & 0 & 0 \\ 1 & 0 & 0 \end{pmatrix} \quad \lambda_{5} = \begin{pmatrix} 0 & 0 & -i \\ 0 & 0 & 0 \\ -i & 0 & 0 \end{pmatrix} \quad \lambda_{6} = \begin{pmatrix} 0 & 0 & 0 \\ 0 & 0 & 1 \\ 0 & 1 & 0 \end{pmatrix}\\
    \lambda_{7} &= \begin{pmatrix} 0 & 0 & 0 \\ 0 & 0 & -i \\ 0 & -i & 0 \end{pmatrix} \quad \lambda_{8} = \frac{1}{\sqrt{3}}\begin{pmatrix} 1 & 0 & 0 \\ 0 & 1 & 0 \\ 0 & 0 & -2 \end{pmatrix}.
\end{align*}
In terms of the identity $I$ and $\{\lambda_{i}\}$, an arbitrary polarization-coherence matrix in 3-D takes the form,
\begin{equation}
    \Phi_{3} = \frac{1}{3}\left(I + \sqrt{3}\sum_{i = 1}^{8}S_{i}\lambda_{i}\right),
\end{equation}
where, $S_{i}$ are the generalized Stokes parameter obtained as $S_{i} = \sqrt{3}\operatorname{Tr}(\lambda_{i}\Phi_{3})/2$. The generalized Stokes parameters are real valued as the polarization-coherence matrix and the Gell-Mann matrices are Hermitian. For $\Phi_{3}$ to be associated with a Jones vector, one must have $\Phi_{3}^{2} = \Phi_{3}$,
\begin{equation} \label{phi2}
\begin{aligned}
    \Phi_{3}^{2} &= \frac{1}{9}\left(I + 2\sqrt{3}\sum_{i}S_{i}\lambda_{i} + 3\sum_{i,j}S_{i}S_{j}\lambda_{i}\lambda_{j}\right)\\
    &= \frac{1}{9}\left(I + 2\sum_{i}S_{i}^{2}I + 2\sqrt{3}\sum_{i}S_{i}\lambda_{i} + 3\sum_{ijk}d_{ijk}S_{i}S_{j}\lambda_{k}\right),
\end{aligned}
\end{equation}
where we have used, $\lambda_{i}\lambda_{j} = \frac{2}{3}\delta_{ij}I + \sum_{k}(d_{ijk} + if_{ijk})\lambda_{k}$ and the antisymmetry of the structure constant $f_{ijk}$. For a pure state,  $\Phi_{3}^{2} = \Phi_{3}$, which implies $\sum_{i}S_{i}^{2} = 1$ and $\sqrt{3}\sum_{ij}d_{ijk}S_{i}S_{j} = S_{k}$. Therefore, fully polarized beams (pure states) lie on the surface of the eight dimensional sphere embedded in the nine dimensional Euclidean space. It is worth noting that, due to the additional condition on the Stokes vectors, only a subset of points on the surface represents the polarization-coherence matrix.

 Hilbert-Schmidt coherence for the polarization-coherence matrix $\Phi_{3}$, in terms of the generalized Stokes parameters is obtained as,
\begin{equation}\label{coh3}
\begin{aligned}
    C_{HS}^{2} &= 2\sum_{i\neq j}|(\Phi_{3})_{ij}|^2\\
     &= \frac{4}{3}\left[S_{1}^{2}+ S_{2}^{2} +S_{4}^{2} +S_{5}^{2}+ S_{6}^{2}+ S_{7}^{2}\right].
\end{aligned}
\end{equation}
The complementary quantity predictability, using (\ref{pred}) is obtained as,
\begin{equation}\label{pred3}
\begin{aligned}
    \mathcal{P}^{2} &= \frac{4}{3}\left[\sum_{i = 1}^{3}((\Phi_{3})_{ii})^{2} - \sum_{i<j}(\Phi_{3})_{ii}(\Phi_{3})_{jj}\right] \\
    &= \frac{4}{3}\left[S_{3}^{2} + S_{8}^{2}\right].
\end{aligned}
\end{equation}
To check explicitly that predictability defined above satisfy the boundary conditions, consider a pure state for a three level system $|\psi\rangle  = a|0\rangle + b|1\rangle + c|2\rangle$, where $a,b,c \in C$ and $|a|^{2} + |b|^{2} + |c|^{2} = 1$. Predictability for this pure state is obtained as,
\begin{equation}
    \mathcal{P}^{2} = \frac{4}{3}\left(|a|^{4} + |b|^{4} + |c|^{4} - |ab|^{2} - |ac|^{2} - |bc|^{2}\right)
\end{equation}
When the probability for finding all the outcomes are equal, i.e., $|a| = |b| = |c| = \frac{1}{\sqrt{3}}$, the predictability is zero and, the coherence is maximum for such states. When the probability of finding one of the outcome is one and others zero, for example, $|a| = 1, |b| = |c| = 0$, the predictability obtained is maximum and coherence for such states is zero. It is worth noting that if $(\alpha, \beta, \gamma)$ represent the probability of the three outcomes respectively, under any permutation of the individual probabilities, the predictability remains unchanged. For pure state, (\ref{coh3}) and (\ref{pred3}) leads to the duality relation,
\begin{equation}
     \mathcal{P}^{2} + C_{HS}^{2} = \frac{4}{3}.
\end{equation}
Complementary character of predictability and coherence for the fields with arbitrary wavefront is evident from the above duality relation. We note in passing, that the degree of polarization for 3-D fields is obtained completely in terms of fraction of energy in the orthogonal components, when the polarization-coherence matrix is diagonal i.e., when coherence is zero. Similar result for the paraxial beam was obtained in Ref. \cite{Qasimi:OL:2007}.

Polarization basis for a non-paraxial beam is three dimensional, with $\mathcal{H}_{p} = \operatorname{Span}\{\textbf{e}_{x}, \textbf{e}_{y}, \textbf{e}_{z}\}$. Along with spatial DOF, it yields an analogous vector space structure to that of a bipartite qutrit system. Therefore, complementary relation for the subsystem of a bipartite qutrit system  will result in the complementary relation for 3D beam given by $\Phi_{3}$. Next, we derive complementary relation for pure and mixed bipartite qutrit systems. 

\subsection*{Complementarity using wedge-product formalism}

Consider a pure two qutrit state given by, $|\psi\rangle_{AB}$, with A and B representing the qutrits. For such systems, the reduced density matrix for the subsystems A and B will be mixed in general, and the duality relation becomes an inequality, $\mathcal{P}^{2} + C_{HS}^{2} \leq \frac{4}{3}$, where the predictability and coherence are obtained for the respective subsystems and, it saturates when the density matrix representing the subsystem is pure. For a pure two qutrit system represented the vectors corresponding to the subsystem $A$ will be $|\phi_{i}\rangle = \langle i_{A}|\psi\rangle$, where $i = 0, 1, 2$. The global entanglement, I-concurrence in terms of the wedge product is given by \cite{bhaskara:QIP:2017},
\begin{equation}\label{entqutrit}
    \mathcal{E}_{AB}^{2} = 4\sum_{i<j}||\phi_{i}\rangle \wedge |\phi_{j}\rangle|^{2}.
\end{equation}
For maximally entangled bipartite qutrit systems, $\mathcal{E}_{AB}^{2} = 4/3$. We observe that for such cases, all the local properties of system are lost, in the sense that coherence and predictability both vanishes. The reduced density matrix of the subsystem A is obtained as, $(\rho_{A})_{ij} = \langle\phi_{j}|\phi_{i}\rangle$. Therefore, the Hilbert-Schmidt coherence for A is obtained as,
\begin{equation}\label{cohqutrit}
  C_{HS}^{2} = 2\sum_{i\neq j}|\langle\phi_{j}|\phi_{i}\rangle|^{2} = 4\sum_{i< j}|\langle\phi_{j}|\phi_{i}\rangle|^{2}.
\end{equation}
From this form it is evident that when the vectors are orthogonal, as in the case of maximally entangled state, the coherence of the subsystem is zero. Since the probability of individual outcomes for A are $\langle\phi_{i}|\phi_{i}\rangle$, from eq. (\ref{pred3}) predictability for this subsystem becomes,
\begin{equation}\label{predqutrit}
    \mathcal{P}^{2} = \frac{4}{3}\left(\sum_{i}|\langle\phi_{i}|\phi_{i}\rangle|^{2} - \sum_{i<j}\langle\phi_{i}|\phi_{i}\rangle\langle\phi_{j}|\phi_{j}\rangle\right),
\end{equation}
It is evident from the above expression that when all the outcomes are equally probable (which is the case for maximally entangled states), the predictability is zero. Therefore, we observe that for the maximally entangled state both the coherence and predictability are zero. Using (\ref{entqutrit}), (\ref{cohqutrit}), and (\ref{predqutrit}), one obtains,
\begin{equation}\label{qutent}
    \begin{aligned}
    \mathcal{E}_{AB}^{2}+ C_{HS}^{2} + \mathcal{P}^{2}  &= 4\sum_{i<j}||\phi_{i}\rangle \wedge |\phi_{j}\rangle|^{2} + 4\sum_{i< j}|\langle\phi_{j}|\phi_{i}\rangle|^{2}\\  
    &+\frac{4}{3}\left( \sum_{i}|\langle\phi_{i}|\phi_{i}\rangle|^{2} - \sum_{i<j}\langle\phi_{i}|\phi_{i}\rangle\langle\phi_{j}|\phi_{j}\rangle\right)\\
    &= \frac{4}{3}\left(\sum_{i}|\langle\phi_{i}|\phi_{i}\rangle|^{2} + 2\sum_{i<j}\langle\phi_{i}|\phi_{i}\rangle\langle\phi_{j}|\phi_{j}\rangle\right)\\ &= \frac{4}{3}\left(\sum_{i}\langle\phi_{i}|\phi_{i}\rangle\right)^{2} = \frac{4}{3}
    \end{aligned}
\end{equation}
where in the second step, Lagrange-Brahmagupta identity \cite{brah}, $|\boldsymbol{a} \wedge \boldsymbol{b}|^{2} = |\boldsymbol{a}|^{2}|\boldsymbol{b}|^{2} - |\boldsymbol{a} \cdot \Bar{\boldsymbol{b}}|^{2}$ and in the last step, normalisation condition is used. Therefore, the global entanglement, predictability and coherence obeys a tight triality relation. When the two qutrit state is separable, entanglement is zero and one obtains the usual duality relation between coherence and predictability. When the state is inseparable, the global entanglement imposes an upper bound on the amount of coherence and predictability the single qutrit can possess. Therefore, local characteristics of the system reduces when the entanglement is non-zero. 

When the two qutrit system itself is mixed, as in the case of two qubit case, entanglement is replaced by the linear entropy of mixedness. For a qutrit with density matrix $\rho_{A}$, the normalized linear entropy of mixedness is given by,
\begin{equation}
    \mathcal{M}^{2}(\rho_{A}) = \frac{3}{2}(1-\operatorname{Tr}(\rho_{A}^{2})).
\end{equation}
Using the expression of $\rho_{A}^{2}$ obtained as in Eq. (\ref{phi2}), one gets,
\begin{equation}\label{triqut}
    \frac{4}{3}\mathcal{M}^{2}(\rho_{A}) + \mathcal{P}^{2} + C_{HS}^{2} = \frac{4}{3}.
\end{equation}
Therefore, in general mixedness of a qutrit bounds the amount of coherence and predictability. Interestingly, for the pure bipartite qutrit system, $\mathcal{E}_{AB}^{2} = \frac{4}{3}\mathcal{M}^{2}(\rho_{A})$, and triality relation (\ref{triqut}) reduces to one with entanglement, coherence and predictability (\ref{qutent}). 
\section{Conclusion}
In conclusion, using the correspondence of paraxial and non-paraxial beams with qubit and qutrit systems, we provided a tight complementarity between coherence, predictability, and entanglement.
Our work is pivotal for non-paraxial optics, emphasizing the saturation of complementarity in mixed states and highlighting the role of entanglement. Furthermore, exploring complex vector space geometry through the wedge-product formalism offers foundational insights, contributing to advancements in utilizing quantum properties. In future, we would like to investigate the role of local entanglement between polarization and field envelope observed in correlated motion of particles in an optical trap \cite{Abanerjee_Panigrahi:PRA:2013} and the role of spin-orbit interaction in near field optics \cite{stafeev:spinorbit:2020}. We expect that the present approach in deriving complementary relations will stimulate further research on classical entanglement. 

\section{Acknowledgement}AKR and NKC thank the Department of Science
and Technology (DST), Govt. of India for Inspire Scholarship. SM acknowledges support from the CSIR project 09/0921(16634)/2023-EMR-I. PKP acknowledges funding from the Department of Science and Technology (DST), India (DST/ICPS/QuEST/Theme-1/2019/6).


\bibliography{sample}
\bibliographystyle{ieeetr}


\end{document}